\documentclass[postprint,3p,times]{elsarticle}

\usepackage[utf8]{inputenc}
\usepackage[T1]{fontenc} 
\usepackage{array} 
\usepackage{hyperref} 
\usepackage{lineno} 
\modulolinenumbers[10] 
\usepackage{nomencl} 
\usepackage{framed} 
\usepackage{multicol} 
\usepackage{etoolbox} 
\usepackage{amsmath} 
\usepackage{graphicx} 
\usepackage{tabularx} 
\newcolumntype{Y}{>{\centering\arraybackslash}X} 
\usepackage{multirow}
\usepackage{enumitem} 
\usepackage{subcaption} 
\usepackage{bm} 
\usepackage{float} 
\usepackage{physics}
\usepackage{pdflscape}
\usepackage{color}
\usepackage{listings}
\journal{SoftwareX}

\begin{document}
\renewcommand{\labelenumii}{\arabic{enumi}.\arabic{enumii}}

\begin{frontmatter}

\title{AlphaPEM: an open-source dynamic 1D physics-based PEM fuel cell model for embedded applications}

\author[FEMTO,LIS]{Raphaël Gass \corref{mycorrespondingauthor}}
\ead{raphael.gass@univ-reunion.fr}
\cortext[mycorrespondingauthor]{Corresponding authors. Website: https://gassraphael.github.io/}

\author[FEMTO]{Zhongliang Li}
\ead{zhongliang.li@univ-fcomte.fr}
\author[LIS]{Rachid Outbib}
\author[FEMTO]{Samir Jemei}
\author[FEMTO,Institut]{Daniel Hissel}

\address[FEMTO]{Université de Franche-Comté, UTBM, CNRS, institut FEMTO-ST, FCLAB, Belfort, France}
\address[LIS]{Aix Marseille Univ, CNRS, LIS, Marseille, France}
\address[Institut]{Institut Universitaire de France, France}

\begin{highlights}
	\item AlphaPEM is a 1D, physics-based, dynamic, biphasic PEM fuel cell system model.
	\item It is designed for embedded application and has a speed-accuracy trade-off.
	\item AlphaPEM is open-source, user-friendly, modular, and coded in Python.
	\item It provides real-time access to the cell's internal states and voltage.
	\item AlphaPEM can automatically calibrate the model's undetermined parameters.
\end{highlights}

\begin{abstract}
	The urgency of the energy transition requires improving the performance and longevity of hydrogen technologies. AlphaPEM is a dynamic one-dimensional (1D) physics-based PEM fuel cell system simulator, programmed in Python and experimentally validated. It offers a good balance between accuracy and execution speed. The modular architecture allows for addition of new features, and it has a user-friendly graphical interface. An automatic calibration method is proposed to match the model to the studied fuel cell. The software provides information on the internal states of the system in response to any current density and can produce polarization and EIS curves. AlphaPEM facilitates the use of a model in embedded conditions, allowing real-time modification of the fuel cell's operating conditions.
\end{abstract}

\begin{keyword}
	Proton exchange membrane fuel cell (PEMFC) \sep Modelling \sep Control-command \sep Automatic calibration
\end{keyword}

\end{frontmatter}

\section*{Metadata}

\begin{table}[!h]
\begin{tabular}{|l|p{6.5cm}|p{6.5cm}|}
\hline
C1 & Current code version & v1.0 \\
\hline
C2 & Permanent link to code/repository used for this code version & https://github.com/gassraphael/AlphaPEM \\
\hline
C3  & Permanent link to Reproducible Capsule & N/A \\
\hline
C4 & Legal Code License & GNU General Public License v3.0 \\
\hline
C5 & Code versioning system used & git \\
\hline
C6 & Software code languages, tools, and services used & Python \\
\hline
C7 & Operating environments & Linux, Mac, Windows \\
\hline
C8 & Link to developer documentation/manual & https://github.com/gassraphael/AlphaPEM \\
\hline
C9 & Support email for questions & raphael.gass@femto-st.fr \\
\hline
\end{tabular}
\caption{Code metadata}
\label{codeMetadata} 
\end{table}

\section{Motivation and significance}

The use of physics-based software simulating PEM fuel cells allows the description of their internal states where sensors cannot be placed, such as the concentration of hydrogen within the catalytic layer of each cell, or the amount of liquid water present in the gas diffusion layer. This information is valuable because fuel cells are complex and difficult converters to operate, and the information provided by non-intrusive sensors do not allow for the precise control of the internal states of fuel cells. Therefore, the use of a model is essential for increasing the accuracy of real-time observation of internal physical states and for deploying specific control based on these observations to enhance the efficiency, power density, and lifetime of fuel cells.  
    
In the current literature, a lack of physics-based PEM fuel cell models that are open to the community is observed. While some commercial software such as COMSOL Multiphysics\textregistered \cite{baoTwodimensionalModelingPolymer2015, mayurMultitimescaleModelingMethodology2015, COMSOLMultiphysicsSoftware}, Ansys Fluent\textregistered \cite{xie3D+1DModelingApproach2020, wuMathematicalModelingTransient2009, fanCharacteristicsPEMFCOperating2017, AnsysFluentFluid}, or Wolfram Mathematica\textregistered \cite{schumacher2+1DModellingPolymer2012, WolframMathematicaModern} allow such modeling, they are not open-source, require expensive licenses and offer limited possibilities for source code modification. The open-source publication of software is, however, a valuable aid to the community, as it not only prevents each research team from having to develop their own simulator from scratch, which is time-consuming, but also improves each software by subjecting it to international critique and allowing for collaborative development, thus enhancing and accelerating research. In this context, a research team from the Institute of Energy and Climate Research, IEK-3, has developed openFuelCell2, an open-source computational fluid dynamics toolbox for simulating fuel cells, based on the open-source library OpenFOAM\textregistered \cite{bealeOpensourceComputationalModel2016, zhangOpenFuelCell2NewComputational2024, zhangOpenFuelCell22023, OpenFOAM2024}. However, all these approaches yield very precise models which are computationally expensive. They are incompatible with embedded applications, which is the objective of this work.

To the authors' knowledge, only two research teams have published PEM fuel cell models for control-command applications as open-source software, both programmed in Matlab. First, Pukrushpan et al. released a 0D dynamic and isothermal model of the fuel cell system in 2004, which includes the auxiliaries and requires very little computational power \cite{pukrushpanControlOrientedModelingAnalysis2004, pukrushpanUniversityMichiganFuel2002}. The aim of this pioneering model was to be used in embedded applications while considering the dynamics of the auxiliaries. However, a physical model that accounts for spatial variations within each component of the fuel cell system would enable more precise diagnosis of the internal states and better support control to optimize these states. Nevertheless, this work has paved the way for the construction of more detailed models. More recently, in 2019, Vetter et al. published a simple and compact software simulating the fuel cell in one-dimensional (1D) steady-state, non-isothermal conditions with two phases of water \cite{vetterFreeOpenReference2019, schumacherPEMFC1DMMM2020}. Although the inclusion of one spatial dimension increases the model's accuracy, the lack of dynamic modeling and consideration of the auxiliaries makes this software incomplete for real-time use in embedded applications. However, it is important to note that this software is primarily intended as a simulation base for more accurate PEM fuel cell models, making it valuable for the community.
    
This paper introduces AlphaPEM, the first open-source, isothermal, two-phase, 1D dynamic model for PEM fuel cell systems. It is designed for real-time model-based diagnosis and control implementation within embedded systems, balancing precision and execution speed. It simulates the dynamic evolution of internal states of the fuel cell, its auxiliaries, and the resultant voltage based on the operating conditions and imposed current density. This software package is written in Python for its readability and ease of writing. It is deployed in open-source with GNU General Public License v3.0 \cite{gnuGNUGeneralPublic2007}. It builds upon the authors' previous works, including a critical review of the underlying physics \cite{gassCriticalReviewProton2024} and an experimentally validated numerical resolution approach \cite{gassAdvanced1DPhysicsbased2025}. An extract of the equations from this multi-physics model is provided in \ref{sec:extract_model_equation}. The modular design of the code allows for easy addition of new features, such as incorporating heat transfer within the fuel cell. Despite the complex physics involved, the code is well-written following the informatics standards \cite{numpyStyleGuide2024} and documented to facilitate its uptake and continuous improvement by the community. AlphaPEM is implemented as a Python class to ease its open-source distribution, leveraging SciPy's classical solver for ordinary differential equations (ODEs). The finite difference problem is solved using SciPy's `solve\_ivp` function \cite{scipyScipyIntegrateSolve_ivp}, employing the implicit 'BDF' method due to the stiff nature of the problem arising from nonlinearities and coupled variables in the ODE system.

The AlphaPEM software package quickly simulates the internal states and voltage dynamics of PEM fuel cell systems for all current densities and operating conditions imposed on it. In particular, it is possible to apply a step current density or use current profiles to generate polarization curves or electrochemical impedance spectroscopy (EIS) curves. The package includes databases from various real fuel cells \cite{fanCharacteristicsPEMFCOperating2017, xieValidationMethodologyPEM2022, EHGroupClean, BalticFuelCells} to facilitate its adoption and allows users to freely insert characteristics of other fuel cells. An automated program for calibrating undetermined parameters is included in AlphaPEM. These parameters are calibrated using the genetic algorithm 'geneticalgorithm2' \cite{pascalGeneticalgorithm22024}, a maintained fork of the widely-used open-source Python program 'geneticalgorithm' \cite{solgiGeneticalgorithm2020}. A graphical user interface is also included to facilitate initial use before delving into the code. Finally, AlphaPEM can be used to compare the results of similar models or assist in the calibration of undetermined parameters in more precise models for which computational time does not allow for accurate calibration within a reasonable timeframe.

\section{Software description}

AlphaPEM is an open-source framework for PEM fuel cell systems modelling, programmed in Python. It is designed for the control and command of embedded systems. Its results reveal the dynamics of the cell internal states and voltage, as well as the balance of plant dynamics, which are vital information for fuel cell management systems.

To use AlphaPEM, it is necessary to install a certain number of packages beforehand.
\begin{lstlisting}[language=Python, numbers=left, frame=single, breaklines=true, tabsize=1]
	git clone https://github.com/gassraphael/AlphaPEM.git # clone the repository
	cd AlphaPEM                              # navigate to the project directory
	python3 -m venv env                   # creation of a new python environment
	source env/bin/activate                      # activation of the environment
	pip install --upgrade pip            # update the Python package manager pip
	pip install numpy scipy matplotlib                       # required packages
	pip install colorama geneticalgorithm2                   # required packages         
	python3 -m pip install git+https://github.com/RedFantom/ttkthemes
\end{lstlisting}

\subsection{Software architecture}

The software architecture of AlphaPEM consists of five directories, each containing several Python files. The root of the software package contains the 'main.py' and 'GUI.py' files. One of these two files must be run to operate the simulator, as they both control the entire software. The 'main.py' file is used for the standard operation of AlphaPEM for programmers. The 'GUI.py' file, which is optional, provides a graphical user interface (GUI) for AlphaPEM to facilitate its use without delving into the program's details. All basic functionalities are included in the GUI without requiring any modifications to other files. However, the GUI does not allow for the calibration of undetermined parameters. The program's results are saved in the '/results' directory.

Next, the directory '/model' contains all the Python files related to the model's physics, such as 'dif\_eq.py', which includes the system of differential equations to be solved. The file 'AlphaPEM.py' contains a class of the same name that represents PEM fuel cell simulators. An object of the AlphaPEM class takes as arguments the set of parameters defining a given fuel cell system, its operating conditions, the imposed current density, and the computing parameters. It returns the evolution of the voltage and all internal states over time. A 'control.py' file is also present, which contains the instructions for dynamically controlling the operating conditions of the fuel cell using the information provided by the model.

The '/modules' directory contains all the Python files that serve as modules for other files. Indeed, to improve the readability of the previous programs, some of the less essential instructions have been written as separate functions and placed in these module files. Each of these module files is named to directly refer to the file it is associated with. For example, 'flows\_modules.py' is used in 'flow.py'. Additionally, a file named 'transitory\_functions.py' is present in this directory and is used in most other programs in the package. It contains a set of mathematical functions that have physical significance for the model, such as the saturation pressure of water vapor.

Finally, the directory '/calibration' contains all the information necessary for calibrating the undetermined parameters of the model. The file 'parameter\_calibration.py' includes the program for performing the calibration, the file 'experimental\_values' contains the experimental information of the fuel cell system that the simulator must represent, the file 'run.sh' contains the instructions to send to the computing cluster to perform the calibration, and the directory '/calibration/results' contains the calibration results.

\begin{figure}[!h]
	\centering
	\includegraphics[width=16.5cm]{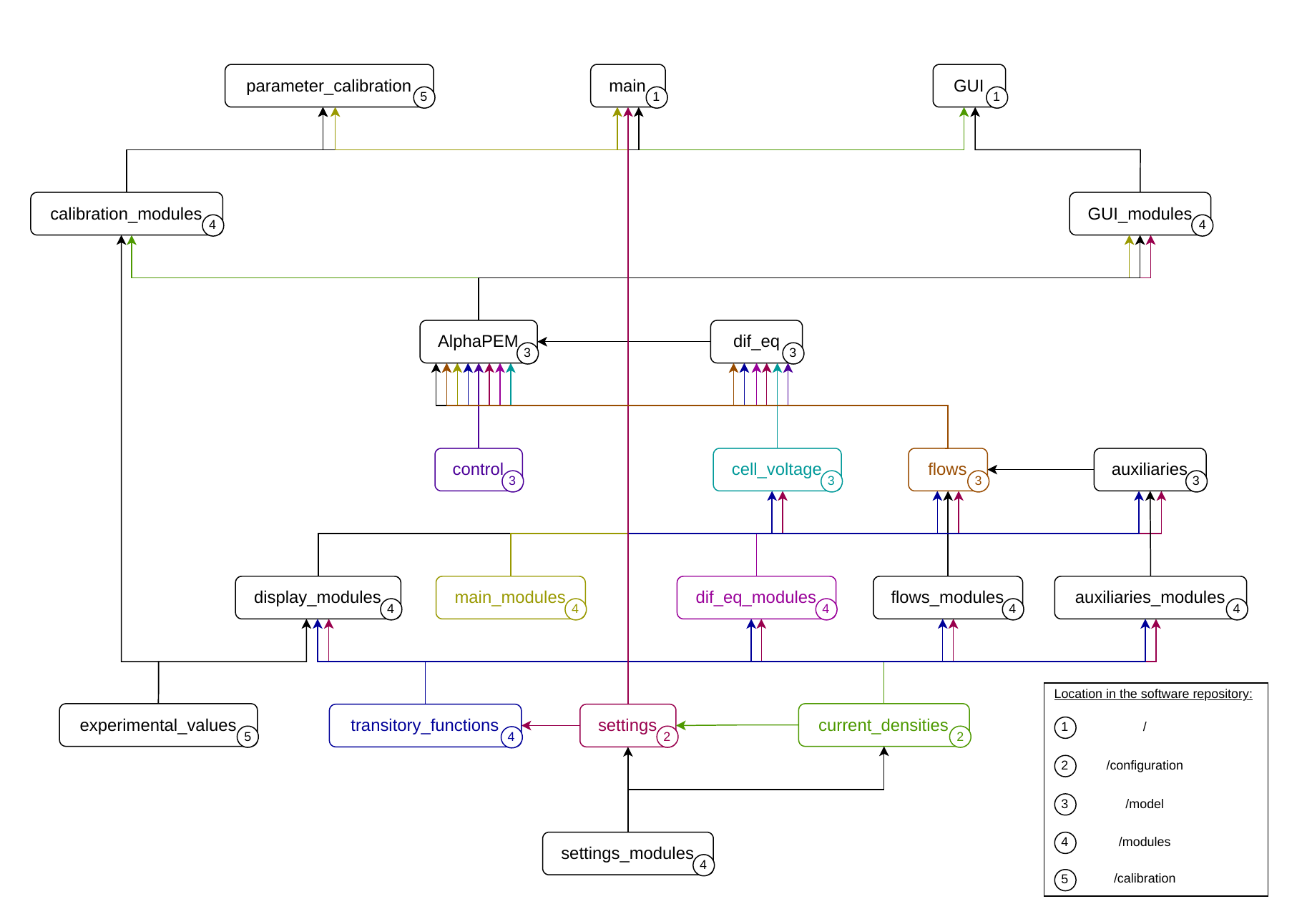}
	\caption{The main structure of AlphaPEM (the reader is referred to the colored web version of this article for improved readability).}
	\label{fig:AlphaPEM_structure}
\end{figure}

Figure \ref{fig:AlphaPEM_structure} represents the structure of AlphaPEM, highlighting the dependencies between the Python files. Each box represents a Python file, with an associated number indicating its location within the software package. An arrow from file A to file B indicates that information from file A are imported into file B. The colors associated with certain boxes and their outgoing arrows improve readability and specifically indicate where these files are imported. This is necessary due to the program's complex overall structure. The boxes that remain black indicate no ambiguity regarding the destination of their arrows. To further enhance readability, the arrows conventionally point from bottom to top or are horizontal. Thus, the files most frequently used by other parts of the program are located towards the bottom of the diagram, while the files executed by the user to start the program are at the top.

\subsection{Software functionalities}
\label{subsec:software_functionalities}

\begin{figure}[!h]
	\centering
	\includegraphics[width=15cm]{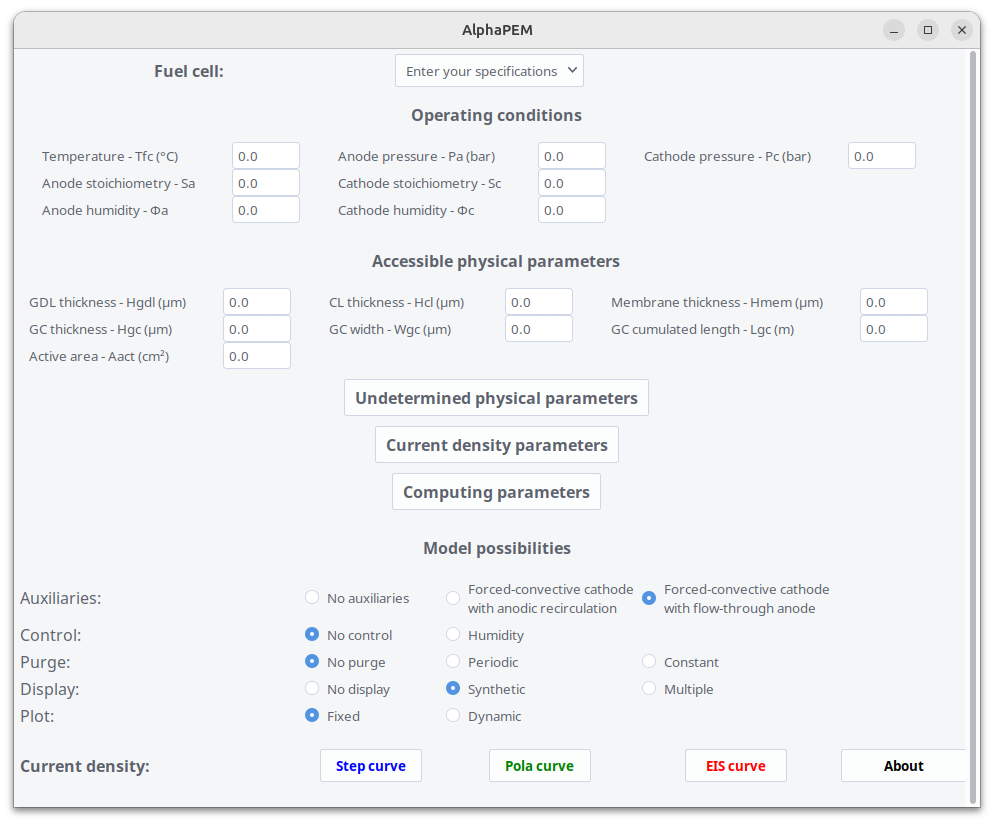}
	\caption{AlphaPEM graphical user interface.}
	\label{fig:demo}
\end{figure}

The usage of the software package AlphaPEM is illustrated by the graphical user interface present in the file '/GUI.py' and displayed in figure \ref{fig:demo}. All the features offered by this GUI are accessible through the files '/main.py' and '/configuration/settings.py'. A fuel cell is characterized by the operating conditions under which it is run (temperature, pressures, stoichiometries and humidities at both anode and cathode), its accessible physical parameters (such as the active area), and its undeterminated physical parameters (such as the tortuosity of the GDL). All these parameters can be adjusted by the user, and predefined configurations based on existing cells can be selected in the 'Fuel cell:' dropdown menu.

Other adjustable parameters exist, here hidden in the GUI to avoid overloading the display. On one hand, the current density parameters allow for the adjustment of the shape of the step current density, or the current density required to create polarization or EIS curves. On the other hand, the computing parameters enable modification of numerical settings, such as the number of points in the numerical model placed in the gas diffusion layer, or the purge times of the stack.

Next, different simulation options can be selected from the 'Model possibilities' menu. It is possible to configure the auxiliaries of the studied fuel cell system as follows: a 'no auxiliaries' system, where matter flows are instantly adjusted to the correct operating conditions at the fuel cell inlet and all gases are evacuated without recirculation; a 'forced-convective cathode with anodic recirculation' system, where the remaining humidified hydrogen at the fuel cell outlet is re-injected at the inlet; or a 'forced-convective cathode with flow-through anode' system, where the anode configuration resembles that of the cathode, with humidified fuel inserted in excess at the inlet and directly evacuated at the outlet. Depending on the selected configuration, the physical modeling of auxiliaries is adjusted, impacting the boundary conditions of the fuel cell system's cells, as detailed in previous work \cite{gassAdvanced1DPhysicsbased2025}. 

Additional options include enabling or disabling control over the operating conditions, choosing the presence or absence of an anode purge, selecting a synthetic or detailed display of results, and displaying results either only at the end of the simulation or with frequent updates during the calculation.

Finally, once these choices are made, the user can generate the model results, which include the internal states (discussed section \ref{subsec:fuel_cell_system_internal_states}) and the voltage of the fuel cell stack, either from a current density step, or a current density producing a polarization curve or an EIS curve. The GUI limits the simulation possibilities to these three types of current densities, but from the source code, it is possible to use any physically acceptable function.

To enable AlphaPEM to simulate a new fuel cell, it is necessary to calibrate certain undetermined physical parameters so that the software results can correspond to the actual cell. Indeed, there are several physical parameters dependent on the specific fuel cell used that are difficult to obtain without the manufacturer's information, such as the porosity of the gas diffusion layers. These parameter values can be approximated using experimental data from the fuel cells. To do this, the user can perform automated calibration of the undetermined parameters using the AlphaPEM Python file '/calibration/parameter\_calibration.py'. This functionality is not available from the GUI. It is necessary to input the experimental values of polarization curves under different operating conditions in the file '/calibration/experimental\_values' (at least three curves), as well as the operating conditions and accessible physical parameters of the studied fuel cell system in the file '/modules/calibration\_modules'.

This automated calibration uses a genetic algorithm. The parameters of this algorithm have been adjusted for this specific optimization problem to achieve a good balance between the accuracy of the calibration and execution speed. These parameters are shown in table \ref{table:genetic_param}. Only the population size and the maximum number of iterations can be modified to match the available computing capacity. It is preferable to have the population size between 100 and 200 individuals and to choose a number that is a multiple of the number of available CPU cores to utilize them fully, as the calculations are parallelized for each member of the same population. The number of iterations should be as large as possible, typically around 1000 to 1500 generations for effective calibration. It is worth noting that the calibration can be resumed from where it previously stopped, allowing multiple computation sessions to finally achieve a satisfactory result. 

Finally, it is preferable to use a computing cluster with many CPU cores for calibration. As an example, the authors successfully performed a calibration with a maximum error of 1.06\% between the experimental and simulated data, after two weeks of calculations on a server equipped with 80 Intel(R) Xeon(R) Gold 6338 CPU cores @ 2.00GHz.

\begin{table*}[!h]
	\centering
	\begin{tabular}{|c|c|} \hline
		
		\textbf{Parameters} & \textbf{Values} \\ \hline
		
		Number of iteration & $1500$ \\
		Population size & $160$ \\
		Mutation probability & $0.33 / nb\_undetermined\_parameters$ \\
		Elit ratio & $1 / population\_size$ \\
		Parents portion & $0.2$ \\
		Crossover type & 'one\_point' \\
		Mutation type & 'uniform\_by\_x' \\
		Selection type & 'roulette' \\ \hline
		
	\end{tabular}
	\caption{Optimised genetic algorithm parameters for AlphaPEM.}
	\label{table:genetic_param}
\end{table*}

\section{Illustrative examples}
\subsection{Polarization curves and parameters calibration}

\begin{figure}[H]
	\centering
	\includegraphics[width=10cm]{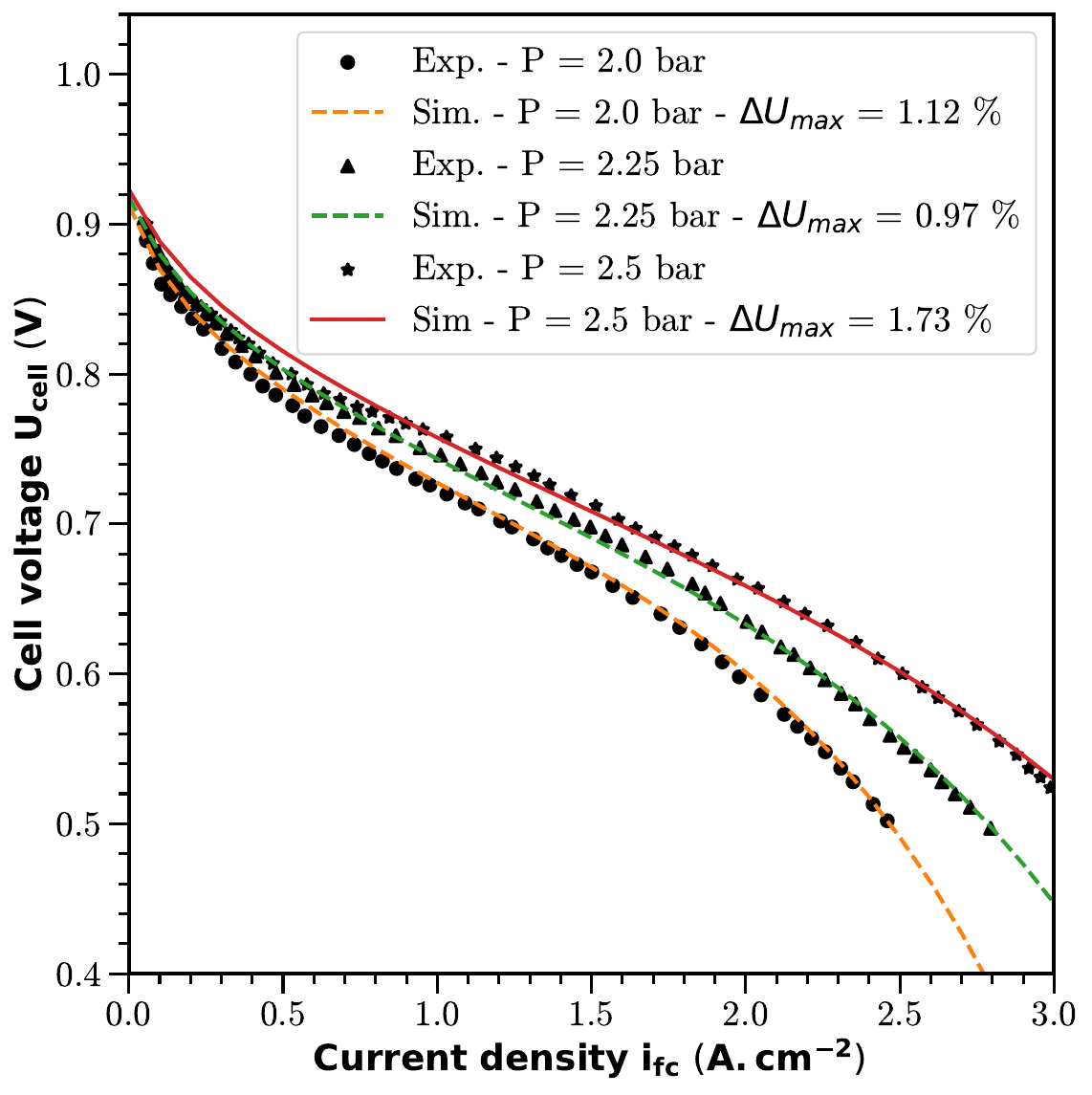}
	\caption{Polarization curves generated by AlphaPEM simulating a single fuel cell system from EH Group \cite{EHGroupClean} under various operating conditions.}
	\label{fig:pola_curve}
\end{figure}

To enable AlphaPEM to simulate the internal states and voltage of a given fuel cell system, it is necessary to provide the simulator with a number of physical parameters to adjust it to the real machine. Some of these parameters are easily accessible, such as the dimensions of each cell, but another part is inaccessible unless the manufacturers have shared them, which is rarely the case. These undetermined physical parameters can be calibrated by AlphaPEM using at least three experimental polarization curves, as discussed section \ref{subsec:software_functionalities}. Once calibration is done, curves like those in figure \ref{fig:pola_curve} can be obtained. The quantity $\Delta U_{max}$ corresponds to the maximum deviation between the experimental and simulated curves. After this calibration, other polarization curves can be simulated by AlphaPEM for all realistic operating conditions.

For an estimation of AlphaPEM's execution speed, the simulator requires less than 35 seconds to simulate one of the polarization curves depicted in figure \ref{fig:step_current} on a mobile workstation equipped with an 11th generation Intel Core i9 processor boasting 16 cores operating at 2.60GHz and 32 GiB of RAM. In this instance, computation occurs at intervals of $0.1$ $A.cm^{-2}$. Thus, optimizing execution speed can be achieved by reducing the precision of the polarization curve.

\subsection{Fuel cell system internal states}
\label{subsec:fuel_cell_system_internal_states}

\begin{figure}[H]
	\centering
	\begin{minipage}{0.75\textwidth}
		\begin{subfigure}[b]{0.49\textwidth}
			\centering
			\includegraphics[width=\textwidth]{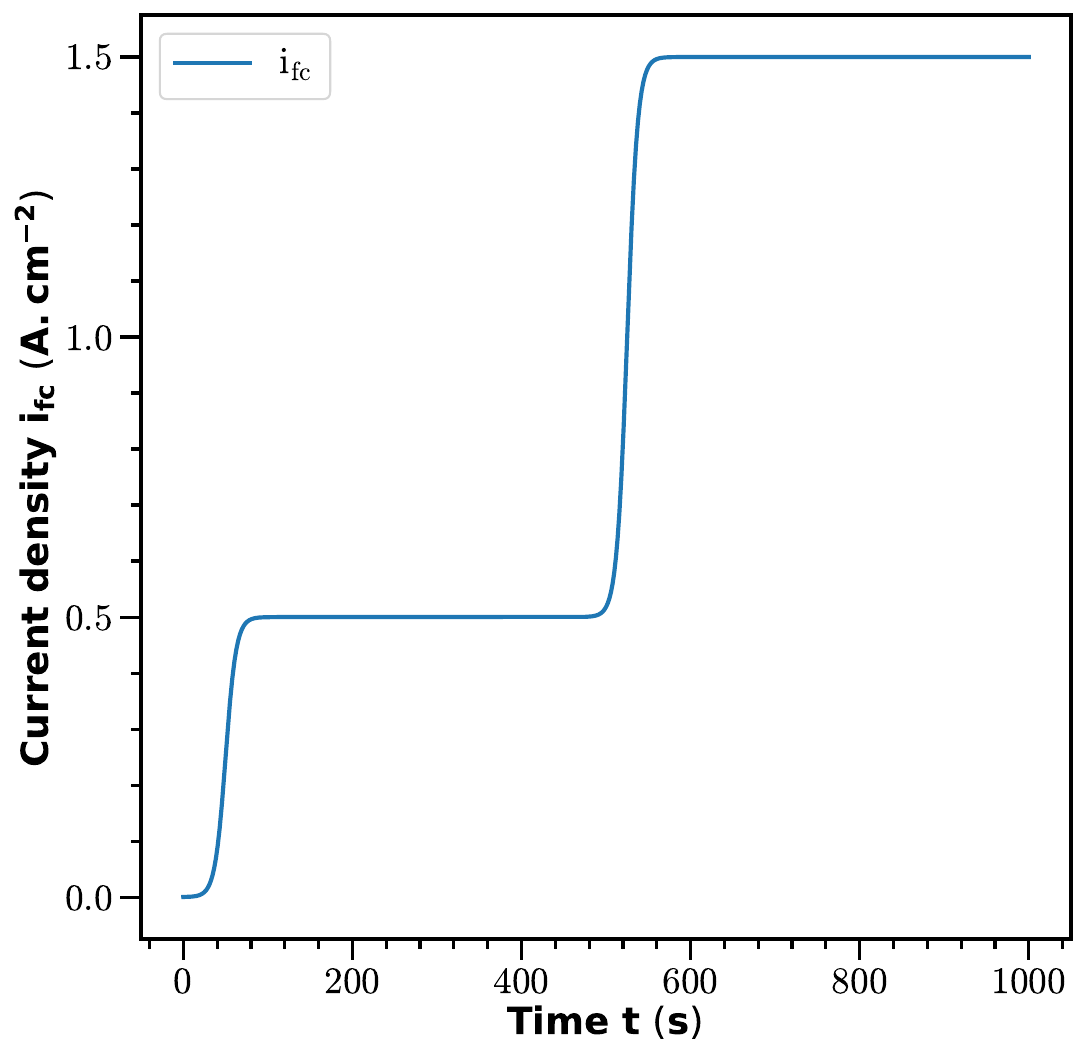}
			\caption{Current density evolution}
			\label{fig:step_current_ifc}
		\end{subfigure}
		\hfill
		\begin{subfigure}[b]{0.49\textwidth}
			\centering
			\includegraphics[width=\textwidth]{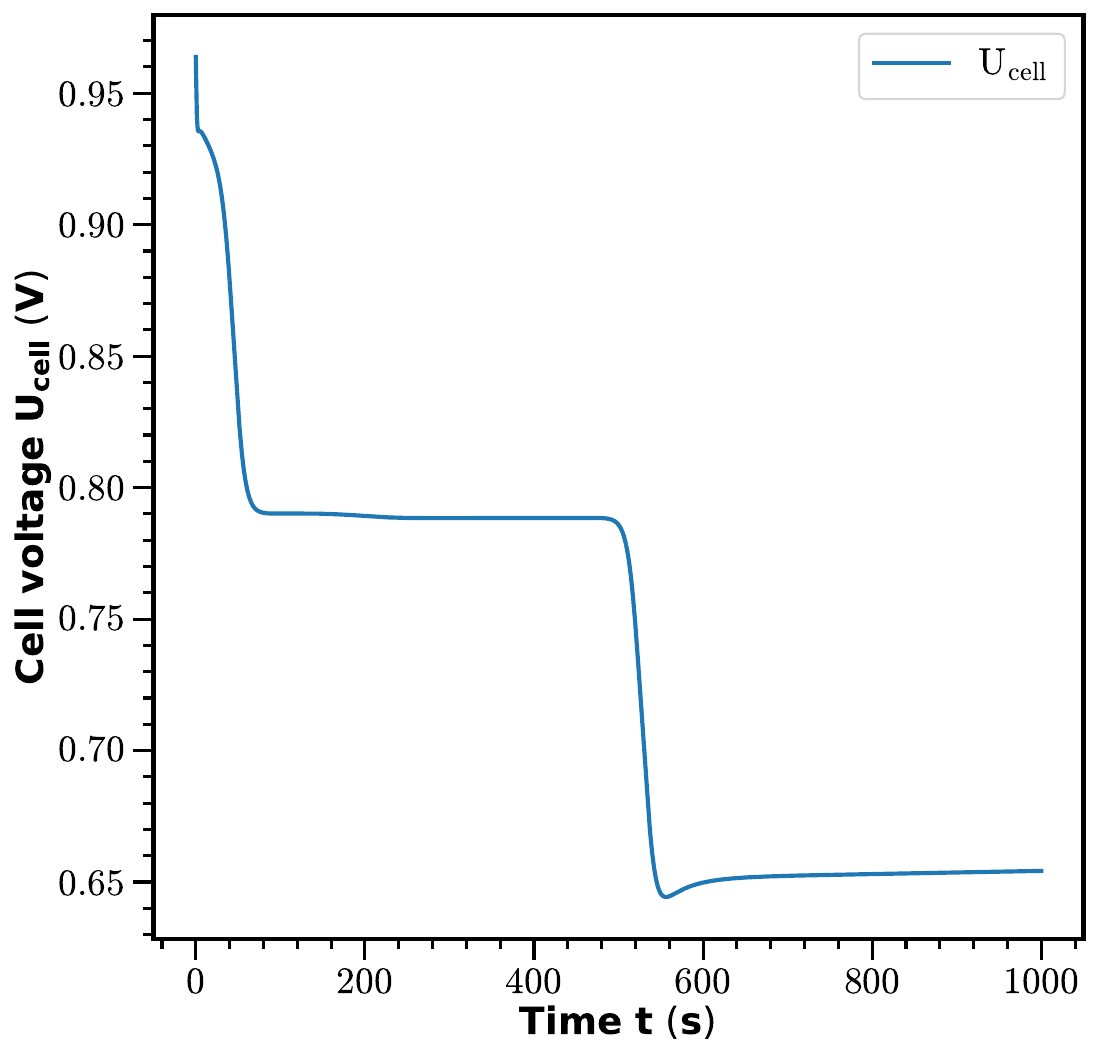}
			\caption{Voltage evolution}
			\label{fig:step_current_Ucell}
		\end{subfigure}
		\vfill
		\begin{subfigure}[b]{0.49\textwidth}
			\centering
			\includegraphics[width=\textwidth]{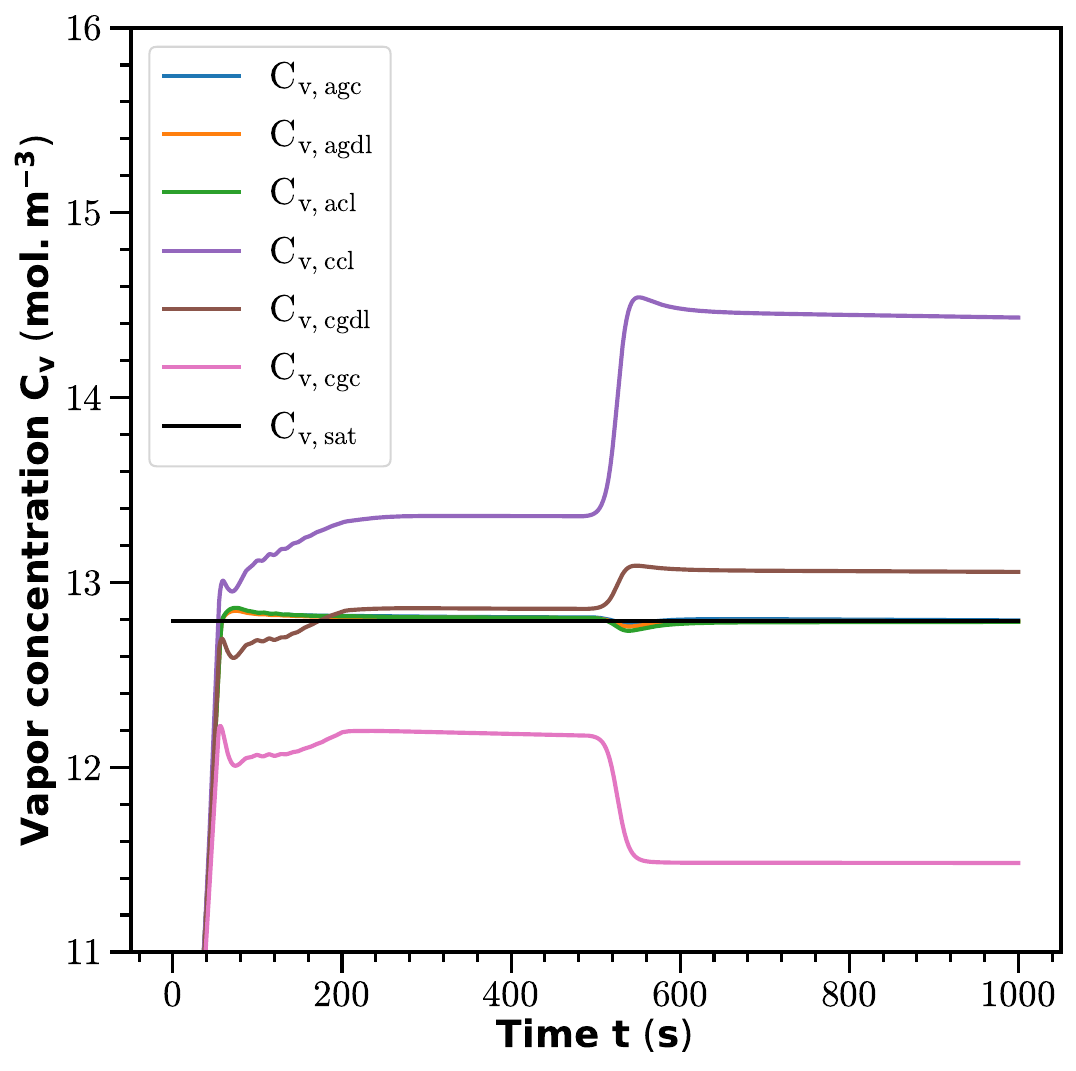}
			\caption{Vapor evolution}
			\label{fig:step_current_Cv}
		\end{subfigure}
		\hfill
		\begin{subfigure}[b]{0.49\textwidth}
			\centering
			\includegraphics[width=\textwidth]{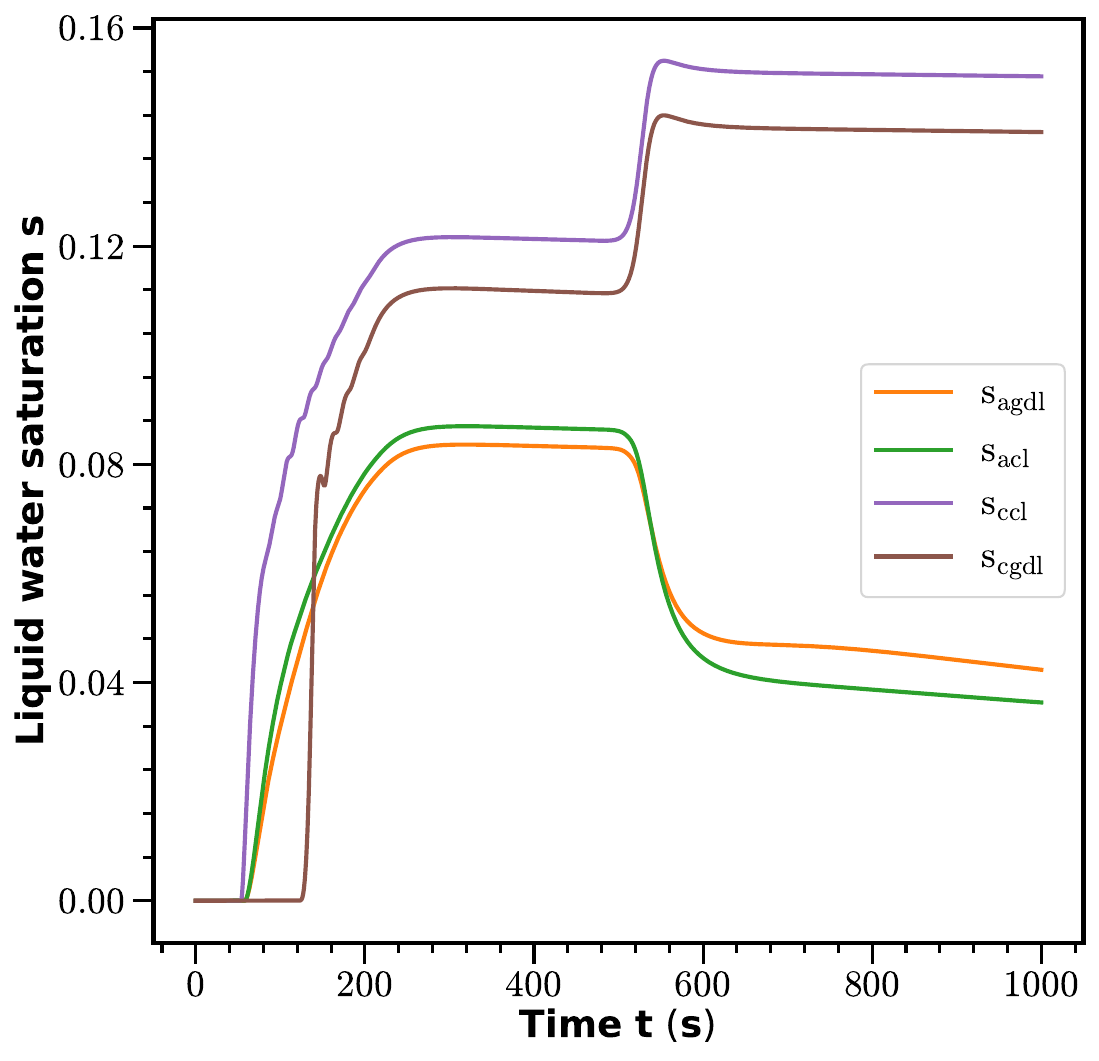}
			\caption{Liquid water evolution}
			\label{fig:step_current_s}
		\end{subfigure}
	\end{minipage}
	\caption{Internal states generated by AlphaPEM simulating a single fuel cell system from EH Group \cite{EHGroupClean} under two current density step stimulations.}
	\label{fig:step_current}
\end{figure}

Once AlphaPEM is calibrated, it is possible to simulate the temporal evolution of the internal states of a given fuel cell. Thus, it is possible to dynamically track at different points within a cell the concentration of water vapor $C_v$, the liquid water saturation $\texttt{s}$, the water content dissolved in the membrane $\lambda$, as well as the concentrations of hydrogen $C_{H_2}$ and oxygen $C_{O_2}$. Similarly, the temporal evolution of cell voltage $U_{cell}$ and various auxiliary variables can be monitored: pressures $P$, relative humidities $\Phi$, mass flows $W$, and throttle areas $A_{bp}$. For instance, figure \ref{fig:step_current} illustrates the temporal evolution of $U_{cell}$, $C_v$ and $\texttt{s}$ for the same fuel cell system previously mentioned in figure \ref{fig:pola_curve}, stimulated by two current density steps. Further curves, details and explanations of the observed evolutions are provided in the authors' previous work \cite{gassAdvanced1DPhysicsbased2025}.

For an estimation of AlphaPEM's execution speed, the simulator requires less than 20 seconds to simulate the 1000-second temporal evolution depicted in figure \ref{fig:step_current} on a mobile workstation equipped with an 11th generation Intel Core i9 processor boasting 16 cores operating at 2.60GHz and 32 GiB of RAM.

\section{Impact}
 
The current energy transition is generating a growing demand for PEM fuel cells with higher power densities and longer lifespans \cite{jiaoDesigningNextGeneration2021, europeanunionCleanHydrogenJoint, amamiyaCurrentTopicsProposed}. In this context, physics-based modeling plays a crucial role in improving the dynamic control of fuel cells, thereby enhancing their performance and longevity \cite{feigaoMultiphysicDynamic1D2010, lunaNonlinearPredictiveControl2016}. However, creating such models is a lengthy and complex task for researchers, as it requires integrating numerous scientific disciplines, including electrochemistry, fluid mechanics, thermodynamics, computational physics, and mathematics. Therefore, it is essential to provide the scientific community with a clear and robust foundational model on which future research can build to advance the field more rapidly. Thus, following the work of Pukrushpan et al. \cite{pukrushpanControlOrientedModelingAnalysis2004} and Vetter et al. \cite{vetterFreeOpenReference2019}, who were the first to disseminate simplified PEM fuel cell models, AlphaPEM is introduced as the first open-source software package modeling dynamically and in 1D fuel cell systems for embedded applications, offering a good compromise between accuracy and execution speed.

AlphaPEM simulates the internal states of fuel cell systems, providing real-time access to information inaccessible by sensors. This is promising for enriching fuel cell control strategies, paving the way for better performance and longevity. Additionally, AlphaPEM is designed with a modular architecture, facilitating the addition of extensions by the community. It is thus easy to incorporate heat transport physics, cell degradation modeling, spatial dimension extension to 1D+1D, and the simulation of cell stacks, to more accurately reflect reality while maintaining sufficient execution speed. These points align with the authors' future research ambitions. The contributions of AlphaPEM thus advance the development of fuel cell systems by facilitating their analysis, control, and improvement.

\section{Conclusions}

The work presents AlphaPEM, an open-source, user-friendly, and modular software package in Python, designed for PEM fuel cell modeling for embedded applications. This framework is based on a 1D finite difference, dynamic, biphasic, and isothermal model of PEM fuel cell systems. It employs a solver using an implicit numerical method to solve the system of differential equations. This model has been experimentally validated in previous studies. In practice, AlphaPEM provides real-time access to the internal states and the voltage of the fuel cell systems and can generate polarization and EIS curves. It can also automatically calibrate the model's undetermined parameters to fit any real fuel cell system. This simulator, therefore, paves the way for improving the real-time control of the operating conditions of fuel cell systems to enhance their performance and longevity. Thermal modeling and spatial extension will be developed in future versions.

\section*{Acknowledgements}
This work has been supported by French National Research Agency via project DEAL (Grant no. ANR-20-CE05-0016-01), the Region Provence-Alpes-Côte d'Azur, the EIPHI Graduate School (contract ANR-17-EURE-0002) and the Region Bourgogne Franche-Comté.

\appendix

\section{Extract of the model equations}
\label{sec:extract_model_equation}

AlphaPEM is based on a multi-physics model with numerous equations that are coupled, nonlinear, and complex to express. These equations have been presented and discussed in the authors' previous work \cite{gassAdvanced1DPhysicsbased2025}, and an extract of them is provided in table \ref{table:extract_flows_equadif}.

\begin{table}[H]
	\centering
	\small
	\linespread{1.5}
	\begin{tabularx}{\linewidth}{|Y|Y|} \hline
		
		\bf{Dynamical models} & \bf{Matter flow expressions} \\ \hline \hline

		\multicolumn{2}{|c|}{\bf{Dissolved water in the membrane}} \\ \hline

		\multirow{2}{*}{
			$\frac{\rho_{mem} \varepsilon_{mc}}{M_{eq}} \frac{d \lambda_{acl}}{d t} = - \frac{J_{\lambda,mem,acl}}{H_{cl}} + S_{sorp,acl} + S_{p,acl}$ \label{subeq:water_content_dynamic_balance_lambda4}} &
		$S_{\text{p},acl} = 2 k_{O_{2}}\left( \lambda_{mem}, T_{fc} \right) \frac{R T_{fc}}{H_{cl} H_{mem}} C_{O_{2},ccl} $ \label{eq:S_prod_acl} \\
		
		&
		$S_{sorp,acl} = \gamma_{sorp}(\lambda_{acl},T_{fc})  \frac{\rho_{\text{mem}}}{M_{\text{eq}}} \left[ \lambda_{\text{eq}}(C_{v,acl},\texttt{s}_{acl},T_{fc})  - \lambda_{acl} \right]$ \label{eq:j_sorp_Ge_a} \\
		
		\multirow{2}{*}{
			$\frac{\rho_{mem}}{M_{eq}} \frac{d \lambda_{mem}}{d t} = \frac{J_{\lambda,mem,acl} - J_{\lambda,mem,ccl}}{H_{mem}}$ \label{subeq:water_content_dynamic_balance_lambda_mem}} &
		$J_{\lambda,mem,acl} = \frac{2.5}{22} \frac{i_{fc}}{F} \lambda_{acl,mem} - \frac{2\rho_{mem}}{M_{eq}} D(\lambda_{acl,mem}) \frac{\lambda_{mem}-\lambda_{acl}}{H_{mem}+H_{cl}}
			\label{eq:water_flow_membrane_acl}$  \\
		
		&
		$J_{\lambda,mem,ccl} = \frac{2.5}{22} \frac{i_{fc}}{F} \lambda_{mem,ccl} - \frac{2\rho_{mem}}{M_{eq}} D(\lambda_{mem,ccl}) \frac{\lambda_{ccl}-\lambda_{mem}}{H_{mem}+H_{cl}}$
			\label{eq:water_flow_membrane_ccl}  \\
		
		\multirow{2}{*}{
			$\frac{\rho_{mem} \varepsilon_{mc}}{M_{eq}} \frac{d \lambda_{ccl}}{d t} = \frac{J_{\lambda,mem,ccl}}{H_{cl}} + S_{sorp,ccl} + S_{p,ccl}$ \label{subeq:water_content_dynamic_balance_lambda_ccl}} &	
		$S_{sorp,ccl} = \gamma_{sorp}(\lambda_{ccl},T_{fc})  \frac{\rho_{\text{mem}}}{M_{\text{eq}}} \left[ \lambda_{\text{eq}}(C_{v,ccl},\texttt{s}_{ccl},T_{fc}) - \lambda_{ccl} \right]$ \label{eq:j_sorp_Ge_c} \\
		
		&
		$S_{\text{p},ccl} = \frac{i_{fc}}{2 F H_{cl}} + k_{H_{2}}\left( \lambda_{mem}, T_{fc} \right) \frac{R T}{H_{cl} H_{mem}} C_{H_{2},acl}$ \label{eq:S_prod_ccl} \\ \hline

	\end{tabularx}
	\caption{Extract of the differential equations and the associated matter transport expressions in the fuel cell derived from the author's previous work \cite{gassAdvanced1DPhysicsbased2025}}
	\label{table:extract_flows_equadif}
\end{table}		

\bibliographystyle{elsarticle-num}
\bibliography{AlphaPEM_software}
\biboptions{sort&compress}

\end{document}